\documentclass[aps,twocolumn,pra,showpacs,floatfix]{revtex4-1}
\usepackage{graphicx}
\usepackage{dcolumn}
\usepackage{amsmath}
\usepackage{color}

\newcommand{\eg}{\emph{e.g.}~}
\newcommand{\ie}{\emph{i.e.}~}
\begin{document}

\title{Limits on Violations of Lorentz Symmetry and the Einstein Equivalence Principle using Radio-Frequency Spectroscopy of Atomic Dysprosium}
\author{M. A. Hohensee}
\email{hohensee@berkeley.edu}
\author{N. Leefer}
\author{D. Budker}
\affiliation{Physics Department, University of California, Berkeley 94720, USA}
\author{C. Harabati}
\author{V. A. Dzuba}
\author{V. V. Flambaum}
\affiliation{School of Physics, University of New South Wales,
Sydney 2052, Australia}
\date{\today}

\begin{abstract}
We report a joint test of local Lorentz invariance and the Einstein equivalence principle for electrons,
using long-term measurements of the transition frequency between two nearly degenerate states of atomic dysprosium.  
We present many-body calculations which demonstrate that the energy splitting of these states is particularly sensitive to violations of both special and general relativity. We limit Lorentz violation for electrons at the level of $10^{-17}$, matching or improving the best laboratory and astrophysical limits by up to a factor of 10, and improve bounds on gravitational redshift anomalies for electrons by 2 orders of magnitude, to $10^{-8}$.  With some enhancements, our experiment may be sensitive to Lorentz violation at the level of $9\times 10^{-20}$.
\end{abstract}
%\pacs{03.30.$+$p, 11.30.Cp, 32.30Jc}

\maketitle

%\paragraph{Introduction}
%\section{Introduction}

Local Lorentz invariance (LLI) and the Einstein equivalence principle (EEP) are fundamental to both the standard model and general relativity~\cite{MTW}.  Nevertheless, these symmetries may be violated at experimentally accessible energy scales due to spontaneous symmetry breaking, or some other mechanism %of an as yet unknown unified theory of physics 
at high energy scales~\cite{Kostelecky:1989,Damour:1996}.  This has motivated the development of many experimental tests of both LLI and EEP~\cite{datatables,Will:2006}, and of a phenomenological framework, known as the standard model extension (SME), which can be used to quantitatively compare these tests' results to one another~\cite{Kostelecky:98}.  This widely used~\cite{datatables} framework augments the standard model Lagrangian with every combination of standard model fields that are not term-by-term invariant under Lorentz transformations, 
while maintaining gauge invariance, energy-momentum conservation, and Lorentz invariance of the total action~\cite{Kostelecky:98}.  Violations of LLI, which themselves constitute violations of EEP~\cite{MTW,Will:2006}, have also been shown to violate other tenets of general relativity%, such as the universality of free fall, and local position invariance
~\cite{KosteleckyTasson2010}.

In this Letter, we show, using many-body calculations, that the energies of two low-lying excited states of dysprosium (Dy)
%the nearly degenerate [Xe]$4f^{10}5d6s$ and [Xe]$4f^{9}5d^{2}6s$ states of dysprosium (Dy)
~\cite{Dzuba86,Dzuba99ab,Dzuba94,Dzuba03} are extremely sensitive to physics that breaks LLI and the EEP in the dynamics of electrons.  We report the results of an analysis of %a modest amount of 
Dy spectroscopy data acquired over two years that significantly improves upon the best laboratory~\cite{Mueller2007} and accelerator~\cite{Altschul:2010a} limits on electron violations of LLI and EEP.  Our result is competitive with some astrophysical bounds~\cite{Altschul:2006}.  We also improve constraints on electron-related gravitational redshift anomalies~\cite{Vessot:1980} by 2 orders of magnitude~\cite{Hohensee2011}.%, to the level of $10^{-8}$~\cite{Hohensee2011}.

The EEP and LLI require that spacetime, while it may be curved, be locally flat, and Lorentzian~\cite{MTW}.  Thus the relative frequencies of any set of clocks at relative rest and located at the same point in (or within a sufficiently small volume of) spacetime must be independent of a) where that point is located in a gravitational potential, and b) the velocity and orientation of their rest frame %relative to any other reference frame 
(LLI).    
In the SME, violation of EEP and LLI for electrons can be described 
by modifying the electron dispersion relation, which in turn causes the energies of bound electronic states to vary with the velocity, orientation, and gravitational potential of their rest frame~\cite{Kostelecky99a,KosteleckyTasson2010}.
%as a modification of the electrons' dispersion relation, which in turn leads to variations in the energies of bound electronic states as a function of velocity and position in an external gravitational potential~\cite{Kostelecky99a,KosteleckyTasson2010}.

We focus on the symmetric, traceless $c_{\mu\nu}$ tensor in the electron sector of the SME, written using coordinates such that the speed of light is a constant $c$ in all frames.  The $c_{\mu\nu}$ tensor modifies the kinetic term in the electronic QED Lagrangian to become~\cite{Kostelecky:98} 
\begin{equation}\label{eq:0}
\mathcal{L}=\tfrac{1}{2}i\bar{\psi}\left(\gamma_{\nu}+c_{\mu\nu}\gamma^{\mu}\right)\stackrel{\leftrightarrow}{D^{\nu}}\psi-\bar{\psi}m\psi,
\end{equation}
where $m$ is the electron mass, $\psi$ is a four-component Dirac spinor, $\gamma^{\nu}$ are the Dirac gamma matrixes, and $\stackrel{\leftrightarrow}{f D^{\nu} g}\,\equiv f D^{\nu} g - g D^{\nu}f$, with $D^{\nu}\equiv \partial^{\nu}-iqA^{\nu}$.  The $c_{\mu\nu}$ tensor is frame dependent~\cite{Kostelecky:98,Kostelecky99a,Bluhm2003,Altschul2010}, and is uniquely specified by its value in a standard reference frame.  We use the Sun-centered, celestial equatorial frame (SCCEF) for this purpose, indicated by the coordinate indexes ($T$,~${X}$,~${Y}$,~${Z}$), for ease of comparison with other results~\cite{datatables}. The component indexes for laboratory frame coordinates are given as $(0,1,2,3)$, where $t=x_{0}/c$ is the time coordinate.  Roman indexes are used to indicate the spatial components of $c_{\mu\nu}$, and are capitalized in the SCCEF frame.  The $c_{\mu\nu}$ tensor has six parity-even components: $c_{TT}$, plus the five $c_{JK}$'s; and three parity-odd components: $c_{TJ}$, which introduce direction and frame dependent anisotropies in the electrons' energy-momentum, or dispersion relation~\cite{Kostelecky:98}.  This shifts the energies of bound electronic states as a function of the states' orientation and alignment in absolute space, breaking both LLI and rotational symmetry~\cite{Kostelecky99a}.  In a gravitational potential, Eq.~\eqref{eq:0} acquires additional terms proportional to $c_{\mu\nu}$ and to the curvature of spacetime~\cite{KosteleckyTasson2010}.  %, and the electron's momentum.  
These arise due to the interplay between the LLI-preserving distortion of spacetime due to gravity, and the LLI-violating distortion due to $c_{\mu\nu}$, generating anomalous gravitational redshifts that scale with the electron's kinetic energy~\cite{KosteleckyTasson2010,Hohensee2011}.

\begin{figure}[t]
\includegraphics[width=8.4cm]{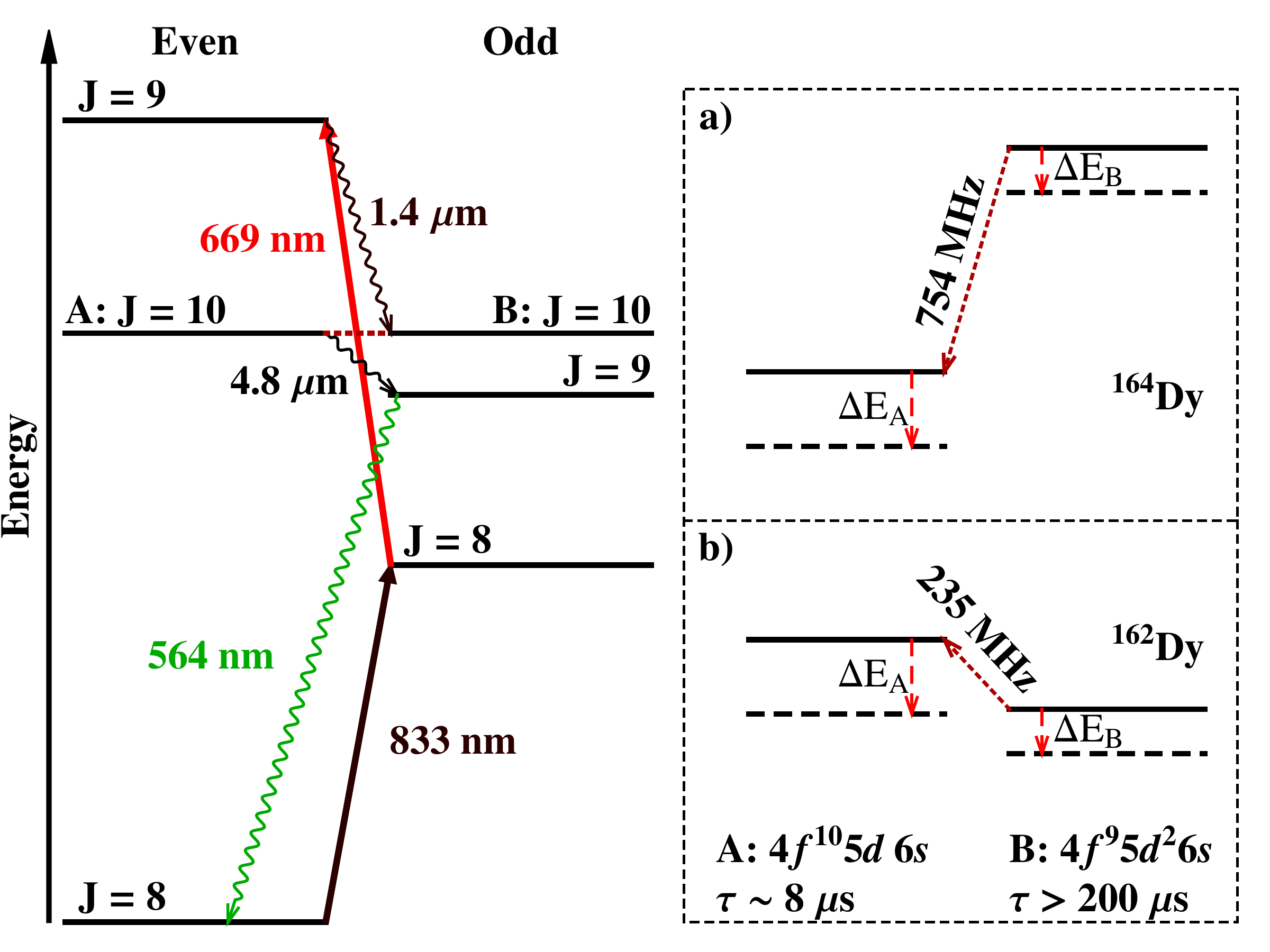}
\caption{\label{fig:diag} Energy levels of Dysprosium.  Atoms are optically pumped (solid lines) to a state which decays (wavy lines) into the metastable state $B$. A linearly polarized rf field drives the $B\rightarrow A$ transition, which is detected via fluorescence at 564nm. Insets a) and b) show the magnified diagram for $^{164}$Dy and $^{162}$Dy, respectively.  Lorentz-symmetry violation shifts the rf resonance by $\delta\omega_{\rm rf}=(\Delta E_{B}-\Delta E_{A})/\hbar$. %Measured frequencies are always positive, so 
The sign of the observed shift depends on the sign of the level splitting.}
\end{figure}

%In general relativity, global Lorentz invariance is broken by the curvature of space-time.  In terms of the global coordinate system, this curvature influences the motion of freely falling particles by modifying their dispersion relations.  General relativity preserves local Lorentz invariance because gravity modifies all particles' dispersion relation in the same way --at any point in space time, it is always possible to define coordinates such that gravity's modifications to particles' dispersion relations drop out.  The additional effect of the  $c_{\mu\nu}$ coefficients, however, makes it impossible to completely remove gravity from the local equations of motion, leading to locally observable gravitational-potential dependent shifts in the energies of bound electronic states, or anomalous gravitational redshift phenomena~\cite{KosteleckyTasson2010,Hohensee2011}.  This energy shift is proportional to the electron's kinetic energy~\cite{Kostelecky99a}.

In terms of spherical-tensor operators, Eq.~\eqref{eq:0} produces a shift $\delta h$ in the effective Hamiltonian for a bound electron with momentum $\vec{p}$ given by~\cite{Kostelecky99a,KosteleckyTasson2010,Kostelecky99b} 
\begin{equation}\label{dh}
\delta h=- \left(C^{(0)}_0-\frac{2U}{3c^{2}}c_{00}\right)\frac{{\vec{p}}^2}{2m} -\sum_{q=-2}^{2}\frac{(-1)^q}{6m} C^{(2)}_q T^{(2)}_{-q} ,
\end{equation}
where we have included the leading order $(2U/3c^{2})c_{00}$ gravitational redshift anomaly~\cite{KosteleckyTasson2010,HohenseeMuellerToBePublished} in terms of the Newtonian potential $U$, and 
\begin{align*}
C^{(0)}_{0}&=c_{00}+\tfrac{2}{3}c_{jj}, \:\:\:\:\:\:\:\:\:\:\:\:\:\:C^{(2)}_{0}=(c_{jj}-3c_{33})\\
 C^{(2)}_{\pm 1}&=\pm6(c_{31}\pm ic_{32}), \:\:\:\:\:\: C^{(2)}_{\pm 2}=3(c_{11}-c_{22}\pm 2ic_{12}),
\end{align*}
are written in terms of the laboratory frame values of the $c_{\mu\nu}$ tensor, with summation implied over like indexes.  Note that $C_{0}^{(2)}$ is also known as $c_{q}$ in the literature~\cite{Kostelecky99a}.  The spherical tensor components of the squared momentum are written as $T^{(2)}_{0}={\vec p\,}^2-3p_3^2$, $T^{(2)}_{\pm 1}=\pm p_3(p_1\pm ip_2)$, and $T^{(2)}_{\pm 2}=(p^2_1-p^2_2)/2\pm ip_1p_2$. The energy shift for a state $|J,M\rangle$ of an atom due to the perturbation (\ref{dh}) is the expectation value of the corresponding $N$ electron operator. Since only tensors with $q=0$ contribute to energy shifts of bound states, we need only calculate matrix elements for the $\vec{p}\,^2$ and $T^{(2)}_0 = {\vec p\,}^2-3p^2_3$ operators. 

Dysprosium, an atom with $66$ protons and a partially filled $f$-shell, is well suited to measuring the electron $c_{\mu\nu}$ coefficients.  It possesses two near-degenerate, low-lying excited states with significant momentum quadrupole moments, opposite parity, and leading configurations: [Xe]$4f^{10}5d6s$, $J = 10$ (state $A$) and [Xe]$4f^{9}5d^{2}6s$, $J = 10$ (state $B$), which differ by a transposition of an electron from the $4f$ to the $5d$ orbital.  The energy difference between these states can be measured directly by driving an electric-dipole transition (Fig.~\ref{fig:diag}) with a radio-frequency (rf) field, and should be particularly sensitive to anomalies proportional to the electrons' kinetic energy, since the $4f$ orbital lies partly within the radius of filled $s$, $p$, and $d$ shells that screen the nuclear charge from the larger $5d$ orbital.  
%The same properties of Dy that make measurements of the $B\rightarrow A$ transition particularly sensitive to variations in the fine structure constant, $\alpha$~\cite{Nguyen2004,Budker07,Leefer2012}, also make them highly sensitive to violations of LLI and EEP. 

To calculate the relevant matrix elements for these states, we use a version of the configuration interaction method optimized for atoms with many electrons in open shells.  This method has been used to calculate energy levels, transition amplitudes, dynamic polarizabilities, ``magic'' frequencies in optical traps, and the effects of $\alpha$ variation and parity violation in Dy and other atoms~\cite{Dzuba08,VN}.  Calculated values of the reduced matrix elements for the $A$ and $B$ states of Dy are presented in Table~\ref{t:matel}, 
and details of their derivation can be found in the Supplemental Material~\cite{supplement}. To check our results, we calculate the fully relativistic matrix elements of $c\gamma^{0}\gamma^{j}p_{j}$ and $ T^{(2)}_0 = c\gamma^{0}(\gamma^{j}p_{j}-3\gamma^{3} p_3)$, corresponding to $\vec{p}^{2}$ and $\vec{p}^{2}-3p_{3}^{2}$~\cite{Kostelecky99b}.  We find good agreement between both calculations, consistent with our initial approximation and intuition.  Measurements of the Dy $B\rightarrow A$ transition are highly sensitive to violations of LLI and EEP because the electrons have more kinetic energy in state $A$ than they do in state $B$.  The same transition is particularly useful for probing variations in the fine structure constant $\alpha$, where it is the energy of state $B$ that depends most strongly on the value of $\alpha$~\cite{Dzuba03,Nguyen2004,Budker07,Leefer2012}.

\begin{table}
\caption{Matrix elements of the relevant operators of Lorentz violation for the states $A$ and $B$ of Dy in units of the Hartree energy $E_{h}=(6.5\times 10^{15}\text{ Hz})h$.}% (in atomic units).}
\label{t:matel}
\begin{ruledtabular}
%                   m a b
  \begin{tabular}{l l c c }
% &   &    \multicolumn{2}{c}{States} \\
 &   &\multicolumn{1}{c}{State A} & \multicolumn{1}{c}{State B} \\
\hline
%\multicolumn{1}{c}{Leading Config.} & & $4f^{10}5d6s$ & \multicolumn{1}{c}{$4f^95d^26s$}  \\
\multicolumn{1}{c}{Term Symbol} &  & \multicolumn{1}{c}{$^3[10]$} & \multicolumn{1}{c}{$^7$H$^o$}  \\
%\multicolumn{1}{c}{J} & & \multicolumn{1}{c}{10} & \multicolumn{1}{c}{10}  \\
 & &\multicolumn{2}{c}{Energies (cm$^{-1}$)} \\
 \multicolumn{1}{c}{Experiment~\cite{Martin}} &  & 19798 & 19798 \\
 \multicolumn{1}{c}{Calculation~\cite{Dzuba2010}} &  & 19786 & 19770 \\
\hline
& & \multicolumn{2}{c}{Matrix Element (units of $E_{h}$)}\\
%\multicolumn{4}{c}{reduced matrix elements (a.u.)} \\
$\langle J\parallel c\gamma^{0}(\gamma^{j}p_{j}-3\gamma^{3} p_3) \parallel J\rangle$  &  & 69.48 & 49.73 \\
  $\langle J\parallel {\vec p}^2-3p_3^2\parallel J\rangle$ &  & 69.84 & 49.89 \\
%\multicolumn{4}{c}{ matrix elements (a.u.)} \\
%  & $ M$ & &  \\
 $\langle JM|{\vec p}^2|JM\rangle$ &  & 437 & 422 \\
\end{tabular}
\end{ruledtabular}
\end{table}

An effusive atomic beam of Dysprosium atoms is produced by a $\sim 1400$~K oven, and is optically pumped into the metastable state $B$ 
%Dysprosium metal is heated in an oven to $\sim1400$~K to produce an effusive atomic beam. The Dy atoms are optically pumped into the metastable state $B$ 
via consecutive laser excitations with 833 nm and 669 nm light, followed by a spontaneous decay. The atoms are resonantly excited from state $B$ to $A$ via an rf electric field, whose linear polarization defines the atoms' quantization axis. The polarization of the excitation laser is chosen to create a symmetric population among the $\pm M$ magnetic sublevels of state $B$ to suppress the effects of Zeeman shifts on our measurement.  Magnetic shielding and Helmholtz coils allow us to cancel background magnetic fields at the level of 20 $\mu$G.  % along this quantization axis. 
The $A$ state relaxes to the ground state in a cascade decay, emitting 564 nm light in the process. The transition frequency is determined by measuring the intensity of the 564 nm fluorescence (with a photomultiplier) as a function of radio frequency, defined relative to a HP5061A Cs frequency reference.  The fractional frequency stability of this reference is rated at better than $10^{-12}$ for $10^{4}$ s of averaging.  We continuously compare the Cs reference to a GPS disciplined Symmetricom TS2700 Rb oscillator, to verify that the fractional drift of the reference is less than $10^{-11}$ over the entire period that data was collected.  Our results depend upon rf measurements with fractional precision larger than $10^{-10}$, and thus we neglect instabilities in the frequency reference in what follows.  
%More details regarding the experimental procedure and apparatus can be found in Refs.~\cite{Budker07,Budker08}.
More details regarding 
%A more detailed description of 
the experimental procedure and apparatus can be found in Refs.~\cite{Budker07,Budker08}.
\begin{figure}[t]
\includegraphics[width=8.4cm]{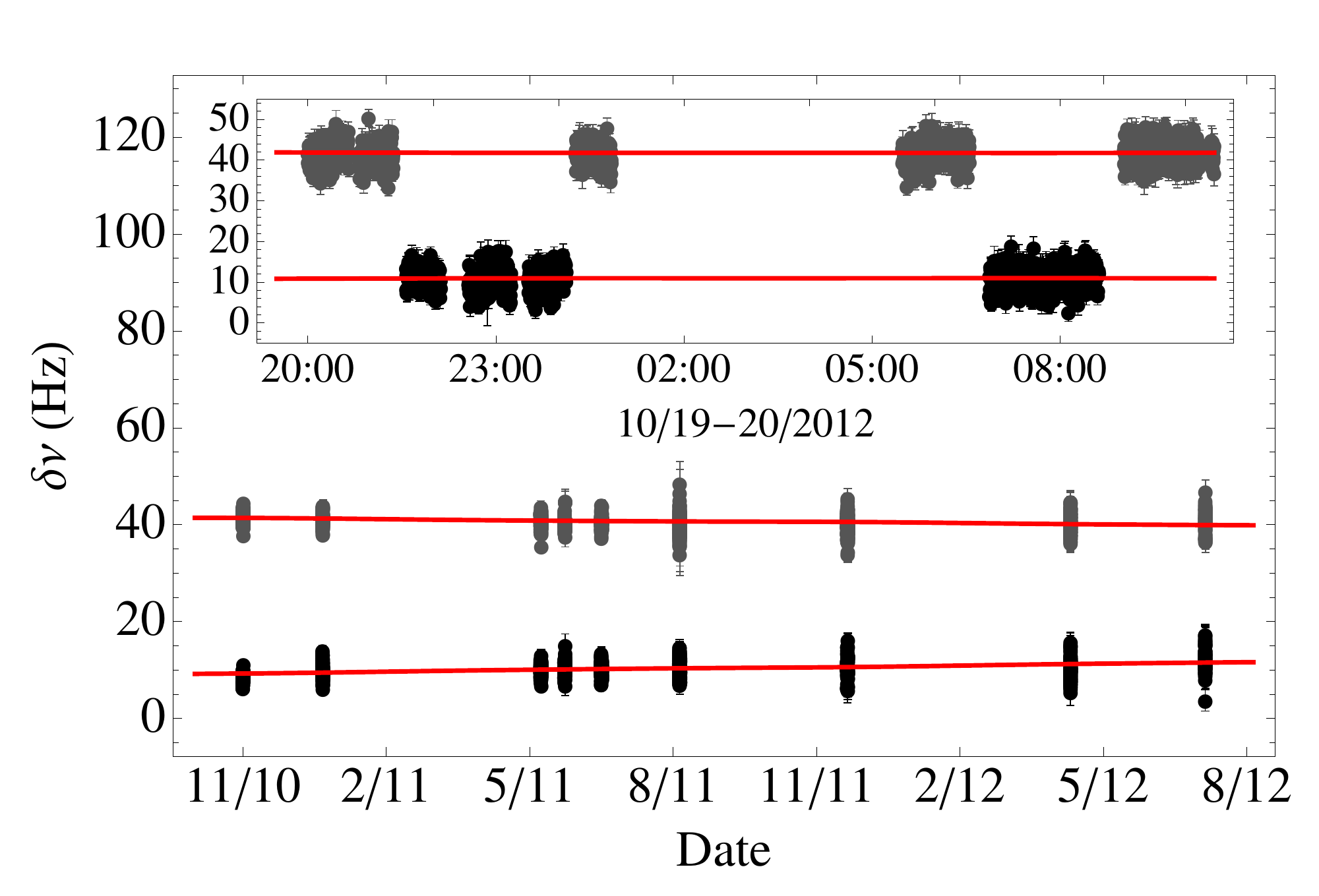}
\caption{\label{fig:1} Full record of frequency measurements for $^{162}$Dy (upper data set) and $^{164}$Dy (lower data set). Frequencies are plotted relative to 234~661~065 Hz for $^{162}$Dy and 753~513~695 Hz for $^{164}$Dy. Error bars are obtained by binning measurements into sets of 20 and calculating the standard error of the mean for each set. The solid line indicates the least-squares fit.  Inset: an expanded view of the most recent measurements beginning on Oct. 19, 2012, with time given in Pacific Standard Time (Coordinated Universal Time minus 8 hours).}
\end{figure} 

We measure the average frequency shift of all populated magnetic sublevels of state $B$ relative to those of state $A$ that are coupled by the rf electric field.  The energy shift of each transition is calculated using Eq.~\eqref{dh} and the calculated reduced matrix elements for each state.  The actual distribution of population among the magnetic sublevels is found by resolving the Zeeman structure of the two states, and measuring the peak amplitude of each transition.  These amplitudes are used as weights in a sum of the shifts of each state due to Eq.~\eqref{dh} to determine the average shift of the unresolved line.  The average shift in the $B\rightarrow A$ transition frequency $\omega_{\rm rf}$ is given by

%We measure the average frequency shift of all populated magnetic sublevels, driven by a linearly polarized rf electric field.  The energy shift of each sublevel is:
%%\begin{multline}\label{lab}
%%\delta E =-\tfrac{1}{2}\left(C_{0}^{(0)}-\tfrac{2U}{3c^{2}}c_{00}\right) \langle {\bf p}^{2}\rangle%\langle JM|{\bf p}^2|JM\rangle  \\
%% -\tfrac{1}{6} C_{0}^{(2)}\tfrac{\langle J\,M\,2\,0 | J\,M\rangle}{\sqrt{2 J +1}} \mathcal{Q},%\langle J\parallel T^{(2)}_{0}\parallel J\rangle,
%%\end{multline}
%\begin{multline}\label{lab}
%\delta E =-\frac{1}{2}\left(C_{0}^{(0)}-\frac{2U}{3c^{2}}c_{00}\right) \langle JM|{\bf p}^2|JM\rangle  \\
% -\frac{1}{6} C_{0}^{(2)}\frac{\langle J\,M\,2\,0 | J\,M\rangle}{\sqrt{2 J +1}} \langle J\parallel T^{(2)}_{0}\parallel J\rangle,
%\end{multline}
%where $J$ is the atom's total angular momentum and $\langle J\,M\,2\,0 | J\,M\rangle$ is a Clebsch-Gordan coefficient.  The actual population distribution among the magnetic sublevels is measured by resolving the Zeeman structure of the two states and measuring the peak amplitude of each transition. These amplitudes are used as weights in an average over the shift for each $|JM\rangle$ state in Eq.~\eqref{lab} to determine the average shift of the unresolved line. %Using the rf-field polarization to define the quantization axis, 
%The average shift in the $B\rightarrow A$ transition frequency $\omega_{\rm rf}$ for $^{162}$Dy and $^{164}$Dy is given by

\begin{equation}
\frac{\delta\omega_{\rm rf}}{2\pi} = \pm\!\left(10^{14}\!\text{ Hz}\right)\!\!\left[ 500\!\left(\!C_{0}^{(0)}\!-\frac{2U_{\odot}}{3c^{2}}c_{00}\!\right)\!+9.1C^{(2)}_{0}\!\right]\!\!,\label{averes}
\end{equation}
where $U_{\odot}=-M_{\odot}G/r_{\rm lab}$ is the Sun's gravitational potential, and $\omega_{\rm{rf}}$ is defined to be positive, producing a positive (negative) shift for $^{164}$Dy ($^{162}$Dy).  This sign difference helps reject background systematics, and is determined by the sign of the energy difference between $A$ and $B$.  The sign of the second term depends on the relative magnetic sublevel populations.

%where $\omega_{\rm{rf}}$ is defined to be positive, producing a positive (negative) shift for $^{164}$Dy ($^{162}$Dy), and $U_{\odot}=-M_{\odot}G/r_{\rm lab}$ is the Sun's gravitational potential. The overall sign of the first term is determined by the sign of the splitting between state $A$ and $B$, and that of the second term is by the populations of the magnetic-sublevels.  The sign flip between the two isotopes helps reject background systematics.
% aids in rejection of % common-mode 
%systematic backgrounds.

The value of $C_{0}^{(0)}$ and $C_{0}^{(2)}$ in the laboratory frame is a function of $c_{\mu\nu}$ in the SCCEF, and the orientation and velocity of the lab. Thus any anomalous $\delta\omega_{\rm rf}$ measured in the lab must vary in time~\cite{Kostelecky99a}.  The precise relation between $C_{0}^{(0)}$ and $C_{0}^{(2)}$ and the SCCEF value of $c_{\mu\nu}$ can be found in the Supplement~\cite{supplement}.  The scalar $c_{TT}$ component of $c_{\mu\nu}$ can be bounded via frame- or gravitational potential-dependent effects, as it contributes to the modulation of $C_{0}^{(2)}$, scaled by Earth's orbital velocity squared $\beta_{\oplus}^{2}\approx 1\times 10^{-8}$, and to that of the larger scalar term in Eq.\eqref{averes} via modulations of the laboratory in the Sun's gravitational potential, which have amplitude $\Delta U_{\odot}/c^{2}= 1.7\times10^{-10}$.
%In the laboratory frame, moving with the Earth's orbital velocity $c\beta_{\oplus}$ relative to the SCCEF, $c_{TT}$ makes a small contribution to $C_{0}^{(2)}$, scaled by a factor of $\beta_{\oplus}^{2}\approx 1\times10^{-8}$.  Over the course of a year, the gravitational potential of the laboratory due to the Sun modulates sinusoidally with amplitude $\Delta U_{\odot}/c^{2}= 1.7\times10^{-10}$, yielding a measurement of $c_{TT}$ via the scalar component of Eq.~\eqref{lab}.  

Using repeated measurements of $\delta\omega_{\rm rf}$ acquired over nearly two years, we obtain constraints on eight of the nine elements of  $c_{\mu\nu}$.% in the SCCEF.  
%The analysis is performed in two parts. 
The %parity-even, 
$c_{JK}$ coefficients are constrained using data collected over the course of 12 h beginning on Oct. 19, 2012.  For each isotope the mean value of 20 successive frequency measurements ($\sim10$ sec) is assigned an error bar according to the standard error of the mean for that bin. The resulting data are fit to Eq.~\eqref{averes} in terms of $c_{JK}$ in the SCCEF~\cite{supplement}, augmented by an independent, constant frequency offset for each isotope.
%The resulting data are fit with Eq.~\eqref{averes}, including an independent, constant frequency offset for each isotope, after a transformation to the SCCEF coordinate system has been made~\cite{supplement}.
  The short duration of this data set allows us to neglect the slow (1 and 2 yr$^{-1}$) variations induced by the $c_{TT}$ and $c_{TJ}$ terms.  These terms are neglected in this fit, as they are suppressed by at least one factor of $\beta_{\oplus}\sim 10^{-4}$, and existing limits~\cite{Altschul:2006} on these terms constrains their contributions well below our statistical sensitivity.
%  are such that their effect is $\sim100$ times less than our statistical sensitivity.
%  
%  and are already sufficiently bounded by astrophysics~\cite{Altschul:2006} that their effect %on our measurements of $c_{JK}$ 
%is roughly $100$ times less than our statistical sensitivity.

\begin{table}[t]
\centering
\caption{Constraints on electron $c_{\mu\nu}$-coefficients from spectroscopy of the rf transitions in $^{162}$Dy and $^{164}$Dy. We use the shorthand notation $c_{X-Y}\equiv c_{XX}-c_{YY}$, $c_{T(Y+Z)} \equiv c_{TY}\cos{\eta} + c_{TZ}\sin{\eta}$, and $c_{T(Y-Z)} \equiv c_{TY}\sin{\eta} - c_{TZ}\cos{\eta}$, where $\eta=23.4^\circ$ is the angle between the Earth's spin and orbital axes. 
Bounds above the horizontal divider are obtained from 12 h of continuous measurement, while those below the line are obtained from analysis of over 2 yr of data, see text.  Some uncertainties for the latter limits are adjusted for systematic error; the statistical uncertainty is then indicated in parenthesis.  
Past bounds on $c_{JK}$, $c_{TJ}$, and $c_{TT}$, and the purely gravitational limit on $c_{TT}$ are from analyses reported in~\cite{Mueller2007},~\cite{Altschul:2006},~\cite{Altschul:2010a}, and~\cite{Hohensee2011}, respectively. } 
\label{table:2}
\begin{ruledtabular}
  \begin{tabular*}{0.47\textwidth}{@{\extracolsep{\fill}}l r r l}
%  Combination & \multicolumn{2}{c}{Limits} & \\
   Combination & \multicolumn{1}{r}{New Limit} & \multicolumn{2}{c}{Existing Limit}  \\
  \hline\noalign{\smallskip}
%  Frame dependent effects\\ \hline\hline\noalign{\smallskip}
  $0.10\,c_{X-Y} -0.99\,c_{XZ}$ & $-9.0\pm11$ & $27\pm19$ & $\times10^{-17}$ \\
%  $0.10\,c_{X-Y} +0.99\,c_{XZ}$ & $-9.0\pm11$ & $-33\pm19$ & $\times10^{-17}$ \\
  $0.99\,c_{X-Y} + 0.10\,c_{XZ}$ & $3.8\pm5.6$ & $-32\pm62$ &$\times10^{-17}$ \\
%  $0.99\,c_{X-Y} - 0.10\,c_{XZ}$ & $3.7\pm5.6$ & $-8.6\pm62$ &$\times10^{-17}$ \\
  $0.94\,c_{XY}-0.35\,c_{YZ}$ & $-0.4\pm2.8$ & $43\pm19$ &$\times10^{-17}$ \\
%  $0.94\,c_{XY}+0.35\,c_{YZ}$ & $-0.4\pm2.8$ & $43\pm19$ &$\times10^{-17}$ \\
  $0.35\,c_{XY}+0.94\,c_{YZ}$ & $3.2\pm7.0$ & $5.3\pm23$ &$\times10^{-17}$ \\\hline\noalign{\smallskip}
%  $0.35\,c_{XY}-0.94\,c_{YZ}$ & $3.2\pm7.0$ & $5.3\pm23$ &$\times10^{-17}$ \\\hline\noalign{\smallskip}
  $0.18\,c_{TX}-0.98\,c_{T(Y+Z)}$ & $0.95\pm18(3.3)$ & $-0.7\pm1.3$ & $\times10^{-15}$ \\
  $0.98\,c_{TX}+0.18\,c_{T(Y+Z)}$ & $5.6\pm7.7(2.4)$ & $-1.4\pm5.4$ & $\times10^{-15}$ \\ %\hline\noalign{\smallskip}
  $c_{T(Y-Z)}$ & $-21\pm19(2.2)$ &$.002\pm.004$ & $\times 10^{-13}$\\
  $c_{TT}$ & $-8.8\pm5.1(4)$ & $10^{-6}(2\pm2)$ & $\times10^{-9}$ \\
%   Grav. potential dependence\\ \hline\hline\noalign{\smallskip}
  $c_{TT}$ (gravitational) & $-14\pm28(9)$ & $4600\pm4600$ & $\times10^{-9}$
%    $0.98\,c_{T(Y+Z)} - 0.19\,c_{TX}$ & $-1.3\pm18(3.3)$ & $0.3\pm1.8$ & $\times10^{-15}$ \\
%  $0.19\,c_{T(Y+Z)} + 0.98\,c_{TX}$ & $7.3\pm6.6(2.4)$ & $-1.5\pm5.4$ & $\times10^{-15}$ \\ %\hline\noalign{\smallskip}
%  $c_{T(Y-Z)}$ & $-38\pm30(5)$ &$.011\pm.028$ & $\times 10^{-13}$\\
%  $c_{TT}$ & $-2.1\pm6.7$ & $10^{-6}(2\pm2)$ & $\times10^{-9}$ \\
%   Grav. potential dependence\\ \hline\hline\noalign{\smallskip}
%  $c_{TT}$ (gravitational) & $-14\pm28(9)$ & $4600\pm4600$ & $\times10^{-9}$
  \end{tabular*}
\end{ruledtabular}
\end{table}

The $c_{TJ}$ and $c_{TT}$ coefficients are constrained using data collected between November 2010 and July 2012. The data are binned and assigned error bars as previously described. Since the above analysis of the 12 h data set provides tight constraints on $c_{JK}$ coefficients, the second fit includes only the $c_{TJ}$ and $c_{TT}$ coefficients. The fit routine is the same as before, adding an independent linear slope for each isotope to account for long-term systematic drifts.   The resulting fit includes a large signal for the combination $c_{T(Y-Z)}\equiv\nolinebreak c_{TY}\sin\eta-\nolinebreak c_{TZ}\cos\eta=(-21\pm 2.2)\times 10^{-13}$, where $\eta$ is the Earth's axial tilt.  As such a signal is inconsistent with existing limits on $c_{T(Y-Z)}$~\cite{Altschul:2010a,Altschul:2006}, we suspect the presence of uncontrolled systematic shifts in $\delta\omega_{\rm rf}$ with characteristic modulation frequencies near $1$ and $2$ day$^{-1}$, and amplitude $~300$ mHz.  These systematics may be due in part to, \eg, magnetic field fluctuations ($\sim50$ mHz), blackbody shifts due to changes in the temperature of the spectroscopy chamber ($\sim60$ mHz)~\cite{LeeferBlackbody}, and changes in electronic offsets ($\sim140$ mHz).  Daily fluctuations in these systematic shifts have less effect on our bounds on $c_{TX}$ and $c_{T(Y+Z)}\equiv c_{TY}\cos\eta+c_{TZ}\sin\eta$, as these are primarily sensitive to the yearly modulation signal produced by the larger scalar component of Eq.~\eqref{averes}~\cite{supplement}.  In the presence of correlated noise, our statistical error bars overstate our measurement's precision; thus, we repeat the least-squares analysis without flipping signs for $^{162}$Dy relative to $^{164}$Dy.  
%using a modified Lorentz violating model that does not reverse sign for $^{162}$Dy relative to $^{164}$Dy. 
This model is insensitive to Lorentz violation, but is sensitive to systematic error.  Where they are larger, the absolute mean of each term in this fit 
replaces the statistical error estimated by the original fit.
%are used to estimate systematic uncertainties for each parameter, while the statistical errors are estimated from the original fit.

We have also analyzed our results as a test of the gravitational redshift for electrons in the Sun's gravitational potential by fitting the long term data to terms proportional to the gravitational potential, neglecting frame dependent effects.  We obtain a purely gravitational limit on the electron's $c_{TT}$ coefficient of $-14\pm 28\times 10^{-9}$.

The data and fits are shown in Fig.~\ref{fig:1}. The fit results are displayed in Table~\ref{table:2} with uncertainties quoted for 68\% confidence limits. %Where systematic uncertainties replace statistical uncertainties, the latter are indicated in parenthesis.  
The reduced chi-squared, $\bar{\chi}^2$, for the short and long time scale fits are 1.2 and 1.8, respectively. The larger $\bar{\chi}^2$ of the long-term fit is likely due to uncontrolled systematics that have not been accounted for in our purely statistical estimation of error bars. To obtain conservative estimates on parameter uncertainties we have scaled the statistical error bars in both fits to provide $\bar{\chi}^{2}=1$.  For the parameters bounded by the long-term fit, even the rescaled statistical limits are smaller than our estimated systematic errors, and so we conservatively conclude that these Lorentz-violating coefficients are at least no larger than our estimated systematic error.
%performed an overall scaling of the statistical error bars in both fits to provide $\bar{\chi}^2 = 1$. 

\begin{figure}[t]
\includegraphics[width=3.2in]{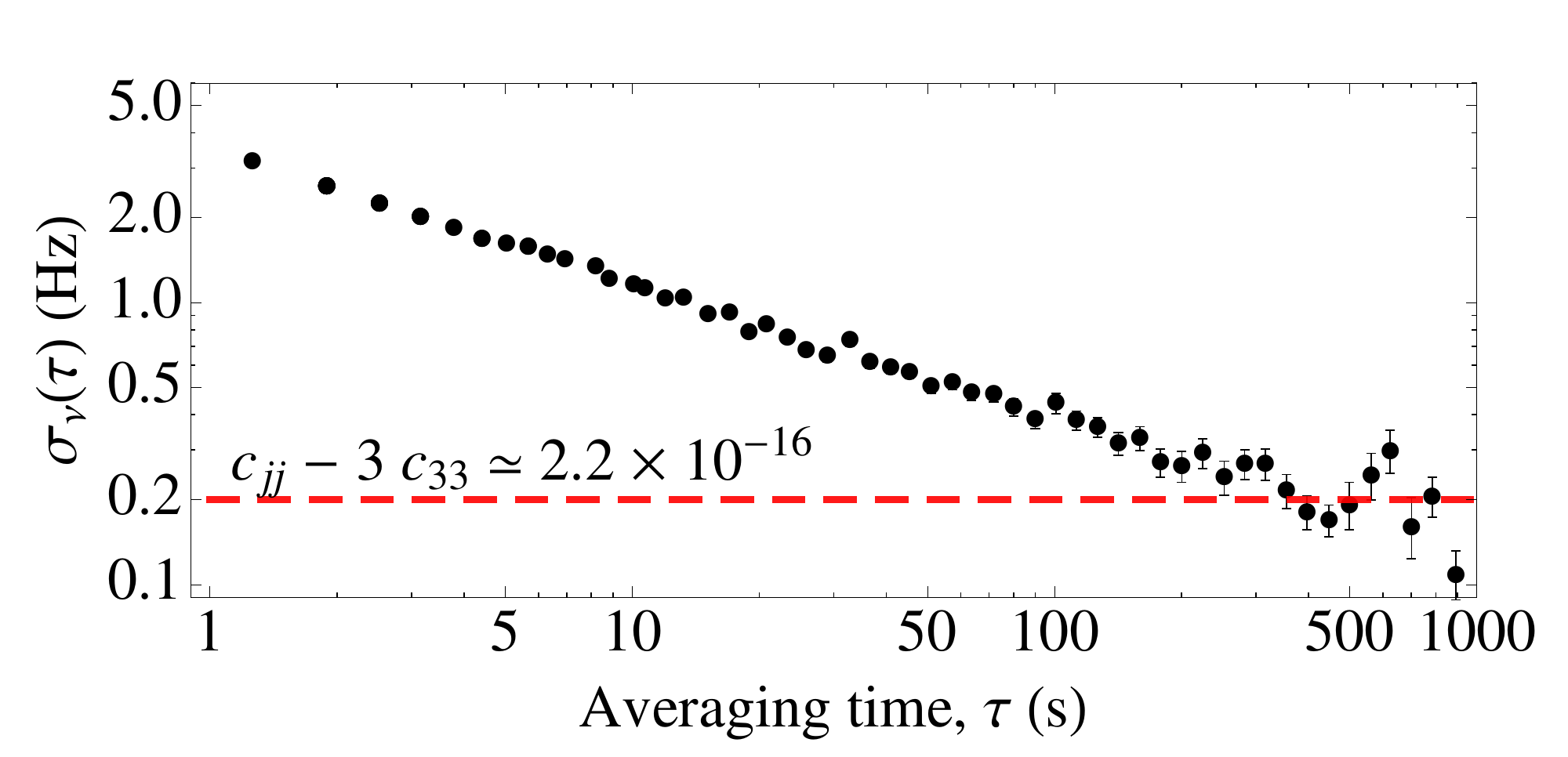}
\caption{\label{fig:2} Allan deviation from a two hour measurement of the $^{164}$Dy transition frequency (7:00 to 9:00 on the Fig.~\ref{fig:1} inset). }
\end{figure}

We have tightened experimental limits on four of the six parity-even components of the $c_{\mu\nu}$ tensor by factors ranging from 2 to 10~\cite{Mueller2007,datatables}.  We report limits on two combinations of the parity-odd $c_{TJ}$ that 
are on par with those set by $\sim 50$ TeV astrophysical phenomena~\cite{Altschul:2006,datatables}.
%are on par with the best astrophysical constraints~\cite{Altschul:2006,datatables}.  
  We improve bounds on electron-related anomalies in the gravitational redshift by a factor of 160, to $2.8\times10^{-8}$.   
With optimization, our experiment could yield significantly improved constraints.  As Fig.~\ref{fig:2} shows, our experiment is statistically sensitive to $C_{0}^{(2)} = c_{jj} - 3c_{33}$ in the lab at the level of $2.2\times10^{-16}$ after 400 sec of averaging. 
At present, we must wait a full day for the Earth to rotate the laboratory in the fixed reference frame, increasing our susceptibility to systematics varying on that time scale.  This could be addressed by active rotation of the entire apparatus, or of the polarization of the rf electric field, making possible statistically limited sensitivities to $C_{0}^{(2)}$ at the level of $1.5\times 10^{-17}$ in one day, and $7.8\times 10^{-19}$ in a year.    
Optically pumping the atoms to the $M = \pm10$ states could increase the experiment's sensitivity to $C_{0}^{(2)}$ by a factor of $\sim4.5$.  
Increasing the interaction time of the atoms in the rf field could gain another factor of two, as the measured linewidth of $40\text{ kHz}$ is twice the natural linewidth of state $A$.  An optimized experiment may thus reach sensitivities at the order of $8.7\times10^{-20}$ in one year.  This would be 3 orders of magnitude better than the presently reported limits on $c_{JK}$, 2 orders of magnitude better than the best sensitivities attainable by existing optical resonator tests~\cite{Herrmann2009Eisele2009}, and could prove more sensitive than astrophysical tests~\cite{Altschul:2006,Klinkhamer:2008a}.  Still narrower linewidths are possible in spectroscopic measurements of the Zeeman and hyperfine structure of the ground state of trapped Dy~\cite{Lev09}, other rare-earth elements, and of the long-lived states of rare-earth ions in doped materials.  Optical transition energies in trapped ion or neutral atom clocks, and of the electronic and rovibrational states of molecules may also be sensitive to $c_{\mu\nu}$.  The latter might also probe violations of LLI and EEP for nuclei.  The derivation of the scalar and quadrupole moments of the involved states will be the subject of future work.

%\section*{Acknowledgments}

We are grateful to Andreas Gerhardus, Paul Hamilton, Alan Kosteleck\'{y}, Jay Tasson, and Christian Weber for stimulating discussions. D.B. acknowledges support from the Miller Institute for Basic Research in Science.  We are grateful to Arman Cing\"{o}z, Valeriy Yashchuk, Alaine Lapierre, and Tuan Nguyen for designing and building the Dy atomic-beam machine.  We thank Holger M\"uller for support of this work.  This work is supported by the Australian Research Council, the National Science Foundation, and the Foundational Questions Institute.
%\clearpage

\section*{Supplemental Material: Matrix Elements}

Here, we present the derivation of the explicit form of the single orbital matrix elements for the operators $\vec{p}^{2}$ and $T^{(2)}_{0}=\vec{p}^{2}-3p_{3}^{2}$.  We also present the matrix elements for the relativistic operators $c\gamma^{0}\gamma^{j}p_{j}$ and $T^{(2)}_{0}=c\gamma^{0}(\gamma^{j}p_{j}-3\gamma^{3}p_{3})$.  

The many-electron state of an atom is a linear combination of Slater determinants, constructed from single-electron orbitals.  These orbitals, in the relativistic limit, can be written as
\begin{equation}
\varphi_{n\kappa m}(\vec{r})=\frac{1}{r}\left(\begin{matrix} f_{n\kappa}\Omega_{\kappa m}(\theta,\phi)\\ i\alpha g_{n\kappa}(r)\Omega_{-\kappa m}(\theta,\phi)\end{matrix}\right),\label{eq:diracspi}
\end{equation}
where $\kappa=\mp(j+1/2)$ (with $j=l\pm1/2$) is the quantum number for the angular momentum of a Dirac spinor, $\alpha$ is the fine structure constant, and the spin-dependent component $\Omega_{\kappa m}(\theta,\phi)$ is given by
\begin{equation}
\Omega_{\kappa m}(\theta,\phi)=\left(\begin{matrix} \pm\sqrt{\frac{\kappa+1/2-m}{2\kappa+1}}Y_{l,m-1/2}(\theta,\phi)\\ \sqrt{\frac{\kappa+1/2+m}{2\kappa+1}}Y_{l,m+1/2}(\theta,\phi)\end{matrix}\right),
\end{equation}
where $Y_{l,m}(\theta,\phi)$ is a spherical harmonic.  The upper component of Eq.~\eqref{eq:diracspi} corresponds to the electron component of the spinor wave function, while the lower component to the positron.  In the non-relativistic limit, we can drop the antimatter component of Eq.~\eqref{eq:diracspi} and evaluate matrix elements in terms of 
\begin{equation}\label{eq:spinor}
\varphi_{n\kappa m}(\vec{r})=\frac{f_{n\kappa}(r)}{r}\Omega_{\kappa m}(\theta,\phi).
\end{equation}

Using the Pauli matrix identity $(\vec{\sigma}\cdot\vec{p})^{2}=\vec{p}^{2}$, with $\vec{\sigma}$ a vector of Pauli matrixes, we can write the action of the operator $\vec{p}^{2}$ on $\varphi_{n\kappa m}(\vec{r})$ as
\begin{equation}
(\vec{\sigma}\cdot\vec{p})^{2}\varphi_{n\kappa m}(\vec{r})=-\frac{\hbar^{2}}{r}\left(\frac{d^{2}f_{n\kappa}}{dr^{2}}-\frac{\kappa(\kappa+1)}{r^{2}}f_{n\kappa}\right)\Omega_{\kappa m}.
\end{equation}
The matrix elements for $\vec{p}^{2}$ are thus equal to $\delta_{\kappa' \kappa}I_{1}$, where $I_{1}$ is the radial integral
\begin{equation}
I_{1}=\hbar^{2}\int_{0}^{\infty} dr\left(\frac{\partial f_{n'\kappa'}}{\partial r}\frac{\partial f_{n\kappa}}{\partial r}+\frac{\kappa(\kappa+1)}{r^{2}}f_{n'\kappa'}f_{n\kappa}\right).
\end{equation}
Similarly, the matrix elements of $p_{3}^{2}=-\hbar^{2}\partial^{2}/\partial z^{2}$ can be found by applying $p_{3}=-i\hbar \partial/\partial z$ to the spinor twice (\emph{e.g.}, by using Eq.~(3.3.3) of~\cite{Szmytkowski2007}).  Using the Wigner-Eckart theorem, the matrix elements of $T^{(2)}_{0}$ can be written in terms of reduced matrix elements after factoring out the $3j$-symbol:
\begin{multline}
\langle n'\kappa' m|T_{0}^{(2)}|n\kappa m\rangle = (-1)^{j'-m}\left(\begin{matrix} j' & 2 & j \\ -m & 0 & m\end{matrix}\right)\\
\times \langle n'\kappa' ||T^{(2)}||n\kappa\rangle.
\end{multline}
These reduced matrix elements are then given by
\begin{equation}
\langle n'\kappa'||T^{(2)}||n\kappa\rangle=\begin{cases}
	A(j',j)I_{1} &\text{if } \kappa'=-\kappa-1\\
	A(j',j)I_{2} &\text{if } \kappa'=-\kappa+1\\
	B(j',j)I_{2} &\text{if } \kappa'=\kappa-2\\
	B(j',j)I_{3} &\text{if } \kappa'=\kappa+2\\
	C(j)I_{1} &\text{if } \kappa'=\kappa
	\end{cases},
\end{equation}
where $A(j',j)$, $B(j',j)$, and $C(j)$ are given by
\begin{equation}
A(j',j)=(-1)^{j_m-j'+1}\left[\frac{6(2j_m+3)(2j_m+1)}{(2j_m+2)2j_m(2j_m+4)}\right]^{1/2}\nonumber
\end{equation}
\begin{equation}
B(j',j)=(-1)^{j_m-j'+1}\left[\frac{3(2j_m+5)(2j_m+3)(2j_m+1)}{2(2j_m+4)(2j_m+2)}\right]^{1/2}\nonumber
\end{equation}
\begin{equation}
C(j)=\left[\frac{(2j+3)(2j+1)(2j-1)}{4j(j+1)}\right]^{1/2}\nonumber
\end{equation}
where $j_m=\mbox{min}(j',j)$.  The radial integrals $I_{2}$ and $I_{3}$ are
\begin{multline}
I_2=\hbar^2\int_0^\infty\text{d}r
\biggl( \frac{\partial f_{n'\kappa '}}{\partial r}
\frac{\partial f_{n\kappa}}{\partial r}-\frac{2\kappa - 1}{r}f_{n'\kappa'}\frac{\partial f_{n\kappa}}{\partial r}\nonumber \\
-\frac{\kappa(\kappa - 2)}{r^2}f_{n'\kappa'}f_{n\kappa}\biggr)\nonumber
\end{multline}
and
\begin{multline}
I_3=\hbar^2\int_0^\infty\text{d}r
\biggl( \frac{\partial f_{n'\kappa '}}{\partial r}
\frac{\partial f_{n\kappa}}{\partial r}+\frac{2\kappa +3}{r}f_{n'\kappa'}\frac{\partial f_{n\kappa}}{\partial r}\nonumber \\
-\frac{(\kappa+3)(\kappa +1)}{r^2}f_{n'\kappa'}f_{n\kappa}\biggr).\nonumber
\end{multline}

As for the the matrix elements of the relativistic form of the operators acting on the four-component spinor in Eq.~\eqref{eq:diracspi}, we find that their angular components $A(j',j)$, $B(j',j)$, and $C(j)$ are the same, while the radial integrals differ, so that
\begin{multline}
\langle n'\kappa'|| c\gamma^{0}(\gamma^{j}p_{j}-3\gamma^{3}p_{3})|n\kappa\rangle\\
=\begin{cases}
	A(j',j)\tilde{I}_{1} &\text{if } \kappa'=-\kappa-1\\
	A(j',j)\tilde{I}_{2} &\text{if } \kappa'=-\kappa+1\\
	B(j',j)\tilde{I}_{3} &\text{if } \kappa'=\kappa-2\\
	B(j',j)\tilde{I}_{4} &\text{if } \kappa'=\kappa+2\\
	C(j)\tilde{I}_{5} &\text{if } \kappa'=\kappa
	\end{cases},
\end{multline}
where
\begin{eqnarray}
\tilde{I}_1&=&-\frac{c \alpha\hbar}{2}  \int_0^\infty\text{d}r
\biggl( (2\kappa -1)g_{n'\kappa'}\frac{\partial f_{n\kappa}}{\partial r}+(2\kappa +3)f_{n'\kappa'}\frac{\partial g_{n\kappa}}{\partial r}\nonumber \\
&&-\frac{(2\kappa-1)(\kappa +1)}{r}g_{n'\kappa'}f_{n\kappa}-\frac{(2\kappa+3)\kappa}{r}f_{n'\kappa'}g_{n\kappa} \biggr),\nonumber\\
%\end{eqnarray}
%\begin{eqnarray}
\tilde{I}_2&=&-\frac{c \alpha\hbar}{2}  \int_0^\infty\text{d}r
\biggl( (2\kappa -3)g_{n'\kappa'}\frac{\partial f_{n\kappa}}{\partial r}+(2\kappa +1)f_{n'\kappa'}\frac{\partial g_{n\kappa}}{\partial r}\nonumber \\
&&+\frac{(2\kappa-3)\kappa}{r}g_{n'\kappa'}f_{n\kappa}+\frac{(2\kappa+1)(\kappa-1)}{r}f_{n'\kappa'}g_{n\kappa} \biggr),\nonumber\\
%\end{eqnarray}
%\begin{eqnarray}
\tilde{I}_3&=&-2c \alpha\hbar  \int_0^\infty\text{d}r
\biggl( f_{n'\kappa'}\frac{\partial g_{n\kappa}}{\partial r}
+\frac{\kappa - 1}{r}f_{n'\kappa'}g_{n\kappa} \biggr),\nonumber\\
%\end{eqnarray}
%\begin{eqnarray}
\tilde{I}_4&=&2c \alpha \hbar \int_0^\infty\text{d}r
\biggl( g_{n'\kappa'}\frac{\partial f_{n\kappa}}{\partial r}
-\frac{\kappa + 1}{r}g_{n'\kappa'}f_{n\kappa} \biggr),\nonumber\\
%\end{eqnarray}
%\begin{eqnarray}
\tilde{I}_5&=&c \alpha \hbar \int_0^\infty\text{d}r
\biggl( g_{n'\kappa'}\frac{\partial f_{n\kappa}}{\partial r}-f_{n'\kappa'}\frac{\partial g_{n\kappa}}{\partial r}\nonumber \\
&&+\frac{\kappa}{r}g_{n'\kappa'}f_{n\kappa}+\frac{\kappa}{r}f_{n'\kappa'}g_{n\kappa} \biggr).\nonumber
\end{eqnarray}

\section*{Supplemental Material: Many-body Calculation}

To evaluate the expectation of the electronic kinetic energy for states $A$ and $B$ in Dy, we use a version of the configuration interaction (CI) method, originally developed for calculating energy levels, electromagnetic amplitudes, dynamics polarizabilities, ``magic'' frequencies in optical traps, the effects of variations in the fine structure constant, and parity violation in atoms with many electrons in open shells~\cite{Dzuba08}.  This method was previously described for the energy levels of Dy in~\cite{Dzuba2010}, and is reviewed here for completeness.  The effective Hamiltonian for $N_{v}$ valence electrons in an atom ($N_{v}=12$ for Dy) has the form
\begin{equation}
\hat{H}^{\rm eff}=\sum_{j=1}^{N_{v}}\hat{h}_{1j}(r_{j})+\sum_{j<k}^{N_{v}}e^{2}/r_{jk},
\end{equation}
where $e$ is the elementary electron charge, and $\hat{h}_{1j}$ is the one-electron Hamiltonian
\begin{equation}
\hat{h}_{1j}(r_{j})=c\vec{\alpha}\cdot\vec{p}+(\beta-1)m_{e}c^{2}-\frac{Ze^{2}}{r}+V_{core}+\delta V,
\end{equation}
where $\alpha_{j}=\gamma^{0}\gamma_{j}$, and $\beta=\gamma^{0}$.  Here, $V_{core}$ is the Hartree-Fock potential due to the core electrons, and $\delta V$ simulates the effects of correlations between core and valence electrons.  It is also known as the polarization potential, and has the form
\begin{equation}\label{eq:deltapol}
\delta V=-\frac{\alpha_{p}}{2(r^{4}+a^{4})},
\end{equation}
where $\alpha_{p}$ is the core polarizability, and $a$ is a cutoff parameter (here, the Bohr radius $a=a_{B}$).
\begin{table} % [h]
\caption{Configurations and effective core polarizabilities
  $\alpha_p$ (a.u.) used in the calculations.}
\label{t:a}
\begin{ruledtabular}
\begin{tabular}{rlll}
%\hline \hline
N & \multicolumn{1}{c}{Parity} &
\multicolumn{1}{c}{Configuration} &
\multicolumn{1}{c}{$\alpha_p$} \\
\hline
 1 & Even & $4f^{10} 6s^2$   &  0.4 \\
 2 & Even & $4f^{10} 6s 5d$  &  0.4006 \\
 3 & Even & $4f^9 6s^2 6p$   &  0.4039 \\
 4 & Even & $4f^9 5d 6s 6p$  &  0.389 \\
 5 & Even & $4f^{10} 6p^2$   &   0.4 \\
 6 & Even & $4f^9 5d^2 6p$   &   0.4 \\
 7 & Odd  & $4f^9 5d^2 6s$   &  0.3947 \\
 8 & Odd  & $4f^9 5d 6s^2$   &  0.3994 \\
 9 & Odd  & $4f^{10} 6s 6p$  &  0.397 \\
10 & Odd  & $4f^{10} 5d 6p$  &  0.4 \\
11 & Odd  & $4f^9 5d 6p^2$  &  0.4 \\
12 & Odd  & $4f^9 6s 6p^2$  &  0.4 \\
%\hline\hline
\end{tabular}
\end{ruledtabular}
\end{table}
Table~\ref{t:a} lists the configurations considered in our calculation.  The self-consistent Hartree-Fock procedure is performed separately for each configuration.  Next, valence states obtained from the Hartree-Fock calculations are used as basis states for the CI calculation.  The CI method requires that the atomic core remain the same for each outer electron configuration.  We select the core state that corresponds to the ground state configuration.  Changes in the core state due to changes in the valence state are small, and can be neglected, as the $6s$, $6p$ and $5d$ shells are comparatively distant, and the potential they produce at the core is a nearly uniform perturbation.  We can thus model the valence potential as shifting the energy of the core eigenstates, without modifying the wave functions.  The $4f$ electrons, on the other hand, lie nearer to the core, and thus have a stronger effect.  In all cases considered here, however, only one of the ten $4f$ electrons ever change state.  Thus their total effect on the atomic core will also be small.  A more detailed discussion of the effects of valence electrons on atomic core states can be found in Refs.~\cite{VN}.

The form of $\delta V$ in Eq.~\eqref{eq:deltapol} is chosen to coincide with the standard polarization potential at large distances $(-\alpha_{p}/2r^{4})$.  We treat the $\alpha_{p}$ for each configuration as a fit parameter, chosen to match experimentally measured energy intervals between states of different configurations.  Their values are displayed in Table~\ref{t:a}.  The values of $\alpha_{p}$ are similar for all configurations considered.  This is expected, as the state of the core electrons is the same for each valence configuration.  The small differences in $\alpha_{p}$ for different configurations help compensate the effect of the incompleteness of our valence electron basis, and other imperfections in the calculation.

\section*{Supplemental Material: Frame Dependence of $\delta\omega_{\rm rf}$}

\begin{table*}[t]
\caption{\label{tab:cjkcoefs}Dominant time-varying terms in the fit for the $c_{JK}$ coefficients.  The frequencies $\omega_{\oplus}$ and $\Omega$ and the sidereal-day and yearly frequencies, respectively. The colatitude of the experiment is given by $\chi \sim 52.1^\circ$, $\theta \sim 15^\circ$ is the angle the quantization axis is rotated towards the North from East, and $\eta \sim 23.4^\circ$ is the angle between the ecliptic and the Earth's equatorial plane. The orbital boost is $\beta_\oplus \sim 10^{-4}$.  The constants $\mathcal{S} \sim \mp5\times10^{16}$~Hz and $\mathcal{Q} \sim \mp 9.1\times10^{14}$~Hz are the scalar and quadrupole shifts, respectively, from Eq.~(3).  For ease of comparison with Table~II, we have defined $c_{X-Y}\equiv c_{XX}-c_{YY}$ and $c_{X+Y}\equiv c_{XX}+c_{YY}$.  Additional terms of $O(\mathcal{Q}\beta_{\oplus}^{2})$ have been suppressed, although they are included in our fits.}
\begin{ruledtabular}
\begin{tabular}{l|cc}
 $\omega_{j}$ & $S_{j}$ & $C_{j}$\\ \hline \\%$\cos\omega T$ & $\sin2\omega T$ & $\cos2\omega T$ \\
 $\omega_{\oplus}$ & $3\mathcal{Q}\left(c_{XZ}\sin\chi\sin2\theta+c_{YZ}\sin2\chi\sin^{2}\theta\right)$ & $3\mathcal{Q}\left(c_{XZ}\sin2\chi\sin^{2}\theta-c_{YZ}\sin\chi\sin2\theta\right)$\\
 $2\omega_{\oplus}$ & $-\tfrac{3}{2}\mathcal{Q}\left[c_{X-Y}\cos\chi\sin2\theta-\tfrac{1}{2}c_{XY}\left(1+3\cos2\theta-2\cos2\chi\sin^{2}\theta\right)\right]$ & $3\mathcal{Q}\left[\begin{matrix}c_{XY}\cos\chi\sin2\theta+ \\ \tfrac{1}{8}c_{X-Y}\left(1+3\cos2\theta-2\cos2\chi\sin^{2}\theta\right)\end{matrix}\right]$\\
 $2\Omega$ & - & $-c_{X+Y}\tfrac{5}{12}\mathcal{S}\beta_{\oplus}^{2}\sin^{2}\eta$
\end{tabular}
\end{ruledtabular}
\end{table*}

The laboratory frame components of the $c_{\mu\nu}$ tensor are found in terms of the sun-centered equatorial frame (SCCEF) components by requiring that the Lagrangian of Eq.~(1) be invariant under the observer Lorentz transformation between frames. This condition leads to

\begin{table*}[ht]
\caption{\label{tab:ctjcoefs}Dominant time-varying terms in the fit for the $c_{TJ}$ and $c_{TT}$ coefficients.  The frequencies $\omega_{\oplus}$ and $\Omega$ and the sidereal-day and yearly frequencies, respectively. The colatitude of the experiment is given by $\chi \sim 52.1^\circ$, $\theta \sim 15^\circ$ is the angle the quantization axis is rotated towards the North from East, and $\eta \sim 23.4^\circ$ is the angle between the ecliptic and the Earth's equatorial plane. The orbital boost is $\beta_\oplus \sim 10^{-4}$.  The constants $\mathcal{S} \sim \mp5\times10^{16}$~Hz and $\mathcal{Q} \sim \mp9.1\times10^{14}$~Hz are the scalar and quadrupole shifts, respectively, from Eq.~(3).  For ease of comparison with Table~II, we have defined $c_{T(Y+Z)}\equiv c_{TY}\cos\eta+c_{TZ}\sin\eta$ and $c_{T(Y-Z)}\equiv c_{TY}\sin\eta-c_{TZ}\cos\eta$.  Additional terms of $O(\mathcal{Q}\beta_{\oplus})$ proportional to $c_{TX}$ and $c_{T(Y+Z)}$ have been suppressed, although they are included in our fits. The gravitational terms appear at $\sin\Omega T$ and $\cos\Omega T$ due to a phase offset $\phi_{\odot}=10.4^{\circ}$ between the oscillation of the boost vector (measured from the vernal equinox) and the oscillation of the Earth in the Solar gravitational potential (with perihelion on Jan. 3rd).  The amplitude of the Earth's modulation in the Solar gravitational potential is $\Delta U/c^{2}\approx 1.7\times 10^{-10}$.}
\begin{ruledtabular}
\begin{tabular}{l|ll}
 $\omega_{j}$ & $S_{j}$ & $C_{j}$\\ \hline \\%$\cos\omega T$ & $\sin2\omega T$ & $\cos2\omega T$ \\
 $\Omega$ & 
 	$\tfrac{10}{3}\mathcal{S}\beta_{\oplus}c_{TX}+\tfrac{2\Delta U}{3c^{2}}\mathcal{S}c_{TT}\sin\phi_{\odot}$ & 
	$\begin{matrix}-\tfrac{10}{3}\mathcal{S}\beta_{\oplus}c_{T(Y+Z)}+\tfrac{2\Delta U}{3c^{2}}\mathcal{S}c_{TT}\cos\phi_{\odot} \\ +\tfrac{3}{8}\mathcal{Q}\beta_{\oplus}c_{T(Y-Z)}\sin2\eta\left(1+3\cos2\theta+6\cos2\chi\sin^{2}\theta\right)\end{matrix}$\\
 $2\Omega$ & 
 	- & 
	$\tfrac{3}{16}\mathcal{Q}\beta_{\oplus}^{2}c_{TT}\sin^{2}\eta\left(1+3\cos2\theta+6\cos2\chi \sin^{2}\theta\right)$\\
%	$\tfrac{1}{8}\mathcal{Q}\beta_{\oplus}^{2}c_{TT}\sin^{2}\eta\left(1+3\cos2\theta+6\cos2\chi \sin^{2}\theta\right)$\\
%
$\omega_{\oplus}-2\Omega$ & 
	$3\mathcal{Q}\beta_{\oplus}^{2}c_{TT}\cos^{3}\tfrac{\eta}{2}\sin\tfrac{\eta}{2}\sin^{2}\theta\sin2\chi$ & 
%	$2\mathcal{Q}\beta_{\oplus}^{2}c_{TT}\cos^{3}\tfrac{\eta}{2}\sin\tfrac{\eta}{2}\sin^{2}\theta\sin2\chi$ & 
	$-3\mathcal{Q}\beta_{\oplus}^{2}c_{TT}\sin\eta\left(1+\cos\eta\right)\sin\chi\sin2\theta$\\
%	$-\tfrac{1}{2}\mathcal{Q}\beta_{\oplus}^{2}c_{TT}\sin\eta\left(1+\cos\eta\right)\sin\chi\sin2\theta$\\
 %
 $\omega_{\oplus}-\Omega$ & 
 	$\tfrac{3}{2}\mathcal{Q}\beta_{\oplus}c_{T(Y-Z)}\left(\cos\eta+\cos2\eta\right)\sin2\chi\sin^{2}\theta$ &  
	$-\tfrac{3}{2}\mathcal{Q}\beta_{\oplus}c_{T(Y-Z)}\left(\cos\eta+\cos2\eta\right)\sin\chi\sin2\theta$ \\
 $\omega_{\oplus}$ &
 	  $\tfrac{3}{4}\mathcal{Q}\beta_{\oplus}^{2}c_{TT}\sin2\eta\sin2\chi\sin^{2}\theta$& 
% 	 - $\tfrac{1}{2}\mathcal{Q}\beta_{\oplus}^{2}c_{TT}\sin2\eta\sin2\chi\sin^{2}\theta$& 
	 $-\tfrac{3}{4}\mathcal{Q}\beta_{\oplus}^{2}c_{TT}\sin2\eta\sin\chi\sin2\theta$\\
%	 $-\tfrac{1}{2}\mathcal{Q}\beta_{\oplus}^{2}c_{TT}\sin2\eta\sin\chi\sin2\theta$\\
%
 $\omega_{\oplus}+\Omega$ & 
 	$-\tfrac{3}{2}\mathcal{Q}\beta_{\oplus}c_{T(Y-Z)}\left(\cos\eta-\cos2\eta\right)\sin2\chi\sin^{2}\theta$ &
	$\tfrac{3}{2}\mathcal{Q}\beta_{\oplus}c_{T(Y-Z)}\left(\cos\eta-\cos2\eta\right)\sin\chi\sin2\theta$\\
 $\omega_{\oplus}+2\Omega$ & 
 	$-\tfrac{3}{8}\mathcal{Q}\beta_{\oplus}^{2}c_{TT}\left(2\sin\eta-\sin2\eta\right)\sin2\chi\sin^{2}\theta$&
% 	$\tfrac{1}{4}\mathcal{Q}\beta_{\oplus}^{2}c_{TT}\left(2\sin\eta-\sin2\eta\right)\sin2\chi\sin^{2}\theta$&
	$\tfrac{3}{8}\mathcal{Q}\beta_{\oplus}^{2}c_{TT}\left(2\sin\eta-\sin2\eta\right)\sin\chi\sin2\theta$ \\
%	$\tfrac{1}{4}\mathcal{Q}\beta_{\oplus}^{2}c_{TT}\left(2\sin\eta-\sin2\eta\right)\sin\chi\sin2\theta$ \\
%
$2\omega_{\oplus}-2\Omega$ & 
	$\tfrac{3}{2}\mathcal{Q}\beta_{\oplus}^{2}c_{TT}\cos^{4}\tfrac{\eta}{2}\cos\chi\sin2\theta$ & 
%	$\mathcal{Q}\beta_{\oplus}^{2}c_{TT}\cos^{4}\tfrac{\eta}{2}\cos\chi\sin2\theta$ & 
	$-\tfrac{3}{8}\mathcal{Q}\beta_{\oplus}^{2}c_{TT}\cos^{4}\tfrac{\eta}{2}\left(1+3\cos2\theta-2\cos2\chi\sin^{2}\theta\right)$\\
%	$-\tfrac{1}{4}\mathcal{Q}\beta_{\oplus}^{2}c_{TT}\cos^{4}\tfrac{\eta}{2}\left(1+3\cos2\theta-2\cos2\chi\sin^{2}\theta\right)$\\
%
$2\omega_{\oplus}-\Omega$ & 
	$\tfrac{3}{4}\mathcal{Q}\beta_{\oplus}c_{T(Y-Z)}\sin\eta\left(1+\cos\eta\right)\cos\chi\sin2\theta$ & 
%	$\tfrac{3}{2}\mathcal{Q}\beta_{\oplus}c_{T(Y-Z)}\sin\eta\left(1+\cos\eta\right)\cos\chi\sin2\theta$ & 
	$\tfrac{3}{16}\mathcal{Q}\beta_{\oplus}c_{T(Y-Z)}\sin\eta\left(1+\cos\eta\right)\left(1+3\cos2\theta-2\cos2\chi\sin^{2}\theta\right)$\\
%	$\tfrac{3}{8}\mathcal{Q}\beta_{\oplus}c_{T(Y-Z)}\sin\eta\left(1+\cos\eta\right)\left(1+3\cos2\theta-2\cos2\chi\sin^{2}\theta\right)$\\
%
$2\omega_{\oplus}$ &
	 $-\tfrac{3}{4}\mathcal{Q}\beta_{\oplus}^{2}c_{TT}\sin^{2}\eta\cos\chi\sin2\theta$ & 
%	 $-\tfrac{1}{2}\mathcal{Q}\beta_{\oplus}^{2}c_{TT}\sin^{2}\eta\cos\chi\sin2\theta$ & 
	 $\tfrac{3}{16}\mathcal{Q}\beta_{\oplus}^{2}c_{TT}\sin^{2}\eta\left(1+3\cos2\theta-2\cos2\chi\sin^{2}\theta\right)$\\
%	 $\tfrac{1}{8}\mathcal{Q}\beta_{\oplus}^{2}c_{TT}\sin^{2}\eta\left(1+3\cos2\theta-2\cos2\chi\sin^{2}\theta\right)$\\
%
$2\omega_{\oplus}+\Omega$ & 
	$\tfrac{3}{4}\mathcal{Q}\beta_{\oplus}c_{T(Y-Z)}\sin\eta(1-\cos\eta)\cos\chi\sin2\theta$ &
%	$3\mathcal{Q}\beta_{\oplus}c_{T(Y-Z)}\sin\eta\sin^{2}\tfrac{\eta}{2}\cos\chi\sin2\theta$ &
	$-\tfrac{3}{16}\mathcal{Q}\beta_{\oplus}c_{T(Y-Z)}\sin\eta(1-\cos\eta)\left(1+3\cos2\theta-2\cos2\chi\sin^{2}\theta\right)$\\
%	$-\tfrac{3}{4}\mathcal{Q}\beta_{\oplus}c_{T(Y-Z)}\sin\eta\sin^{2}\tfrac{\eta}{2}\left(1+3\cos2\theta-2\cos2\chi\sin^{2}\theta\right)$\\
%
$2\omega_{\oplus}+2\Omega$ &
	$\tfrac{3}{2}\mathcal{Q}\beta_{\oplus}^{2}c_{TT}\sin^{4}\tfrac{\eta}{2}\cos\chi\sin2\theta$ &
%	$\mathcal{Q}\beta_{\oplus}^{2}c_{TT}\sin^{4}\tfrac{\eta}{2}\cos\chi\sin2\theta$ &
	$-\tfrac{3}{8}\mathcal{Q}\beta_{\oplus}^{2}c_{TT}\sin^{4}\tfrac{\eta}{2}\left(1+3\cos2\theta-2\cos2\chi\sin^{2}\theta\right)$\\
%	$-\tfrac{1}{4}\mathcal{Q}\beta_{\oplus}^{2}c_{TT}\sin^{4}\tfrac{\eta}{2}\left(1+3\cos2\theta-2\cos2\chi\sin^{2}\theta\right)$\\
\end{tabular}
\end{ruledtabular}
\end{table*}

\begin{equation}
c_{\mu\nu} = c_{MN} \Lambda^{M}_\mu\Lambda^{N}_\nu,
\end{equation}
where $c_{M N}$ is the $c$ tensor defined in the SCCEF. In the SCCEF the $Z$ axis points along the Earth's rotation axis, the $X-Y$ plane is the Earth's equatorial plane, and the $X$ axis points from the Earth to the Sun at the vernal equinox. The Lorentz transformation $\Lambda^{M}_\mu$ is a rotation, dependent on experimental geometry and the rotation of the Earth, to align the laboratory defined axes with these SCCEF axes followed by a boost, determined mainly by the Earth's orbital velocity, to the rest frame of the Sun. Our laboratory frame is defined such that $z$ is parallel to the quantization axis defined by the polarization of the rf field.  Thus $z$ points $\theta=15^{\circ}$  north of east, $x$ points $\theta=15^{\circ}$ east of south, and $y$ points vertically downward.  The rotation from the SCCEF to the laboratory frame is given by
\begin{widetext}
\begin{equation}
R=\left(\begin{matrix} 
\cos\chi\cos\theta\cos\omega T_{\oplus}-\sin\theta\sin\omega T_{\oplus} & \sin\theta\cos\omega T_{\oplus} +\cos\chi\cos\theta\sin\omega T_{\oplus} & -\sin\chi\cos\theta \\
-\sin\chi\cos\omega T_{\oplus} & -\sin\chi\sin\omega T_{\oplus} & -\cos\chi \\
-\cos\chi\sin\theta\cos\omega T_{\oplus} -\cos\theta\sin\omega T_{\oplus} & \cos\theta\cos\omega T_{\oplus} -\cos\chi\sin\theta\sin\omega T_{\oplus} & \sin\chi\sin\theta
\end{matrix}\right),
\end{equation}
\end{widetext}
where $\chi=52.1^{\circ}$ is the colatitude of our laboratory, $\omega$ is the angular frequency of the Earth's rotation in the SCCEF (\ie $2\pi\times$ an inverse sidereal day), and $T_{\oplus}$ is measured from the first time that East, as measured in the laboratory, and the SCCEF X-axis coincides after a vernal equinox.  The boost of the laboratory in the SCCEF frame is given by~\cite{datatables}
\begin{equation}
\vec{\beta}=\left(\begin{matrix}
\beta_{\oplus}\sin\Omega T-\beta_{L}\sin\chi\sin\omega T\\
-\beta_{\oplus}\cos\eta\cos\Omega T+\beta_{L}\sin\chi\cos\omega T\\
-\beta_{\oplus}\sin\eta\cos\Omega T
\end{matrix}\right),
\end{equation}
where $\eta=23.4^{\circ}$ is the angle between the ecliptic plane and the Earth's equatorial plane, $\beta_{\oplus}=2\pi(1\text{ a.u.})/c(1\text{ yr})\simeq 10^{-4}$ is the boost from the Earth's orbital velocity, $\beta_{L}=R_{\oplus}\omega\simeq 1.5\times 10^{-6}$ is the boost from the rotational velocity of a point on the Earth's equator, $2\pi/\Omega$ is a sidereal year, and $T$ is the time since the epoch, chosen to be the vernal equinox in the year 2000~\cite{datatables}.  The boost $\beta_{L}\sin\chi$, though included in our fits, is too small to make a significant contribution to our fits, and will henceforth be dropped from this discussion.

Since the transformation between frames is time-dependent, constant values of $c_{MN}$ in the SCCEF give rise to time varying frequency shifts in the laboratory value of $c_{\mu\nu}$, and thus to time variations in $\delta \omega_{\rm rf}$, via Eq.~(3).  Since $c_{MN}$ also gives rise to an anomalous gravitational redshift, there is an additional contribution proportional to $c_{TT}$, such that $\delta \omega_{\rm rf} \approx \mp ( 5\times10^{16}$ Hz$)\times 2\Delta U/(3c^2)\cos(\Omega T-\phi_{\odot})$, as given in Eq.~(3),  where $\Delta U\sim 1.7\times 10^{-10}c^{2}$ is the amplitude of the Earth's yearly modulation in the solar gravitational potential due to the eccentricity of the Earth's orbit, and $\phi_{\odot}$ is such that $\cos(\Omega T -\phi_{\odot})$ is minimized at perihelion ($\sim$Jan 3).

The $c_{JK}$ coefficients contribute leading order energy shifts at much shorter time scales (daily) than the $c_{TJ}$ and $c_{TT}$ coefficients (yearly). As such, the $c_{JK}$ coefficients are constrained using a single long data set acquired over the course of one day to minimize the influence of systematic effects acting on long time scales.  To fit the $c_{JK}$ terms, we assume that the parity-odd $c_{TJ}$ coefficients are as constrained by astrophysical observations~\cite{Altschul:2006}, and that the contribution of the $c_{TT}$ term to the daily modulation of $\delta\omega_{\rm rf}$ is negligible. Although we retain all terms up to $O(\beta_{\oplus}^{2})$ in our fit function, the dominant terms that are relevant to our fit are 
\begin{equation}
\delta\omega_{\rm rf}=A+\sum_{j}\left(S_{j}\sin\omega_{j}T+C_{j}\cos\omega_{j}T\right),\label{eq:fitcjk}
\end{equation}
where $A$ is an independent offset parameter for each isotope, and the relevant frequencies $\omega_{j}$, and quadratures $S_{j}$ and $C_{j}$ which contain the relevant components of $c_{JK}$, are summarized in Table~\ref{tab:cjkcoefs}.  Since our experiment alternates between measuring $\delta\omega_{\rm rf}$ for $^{162}$Dy and $^{164}$Dy, we do not directly compare $\delta\omega_{\rm rf}$ for the two isotopes.  We therefore perform a simultaneous fit of the separate functions~\eqref{eq:fitcjk} for each isotope, subject to the constraint that the values of $c_{JK}$ must be the same for both.

The $c_{TJ}$ and $c_{TT}$ coefficients are constrained using data acquired at irregular intervals over two years.  To fit the $c_{TJ}$ and $c_{TT}$ terms, we set $c_{JK}$ to zero, and drop terms proportional to $\beta_{\oplus}^{2}$ that do not multiply $c_{TT}$, since existing bounds on $c_{X+Y}\equiv c_{XX}+c_{YY}$ are sufficient to ensure its doubly boost-suppressed contribution to $\delta\omega_{\rm rf}$ is negligible at our level of precision~\cite{datatables}.  The fit function is
\begin{equation}
\delta\omega_{\rm rf}=A+M T+\sum_{j}\left(S_{j}\sin\omega_{j}T+C_{j}\cos\omega_{j}T\right),\label{eq:fitctj}
\end{equation}
where as before, $A$ is an isotope-dependent offset, and the $M T$ term is applied to remove any linear drifts.  The relevant frequencies $\omega_{j}$, and quadratures $S_{j}$ and $C_{j}$, which contain the relevant components of $c_{TJ}$ and $c_{TT}$, are summarized in Table~\ref{tab:ctjcoefs}.  We perform a joint fit on the $^{162}$Dy and $^{164}$Dy data as before, and note an apparently large signal for $c_{T(Y-Z)}\approx (38\pm5)\times 10^{-13}$, roughly $7.6$ times the statistical error bar.  As such a result would be inconsistent with other experiments~\cite{datatables}, we suspect it may be due to modulated systematic errors that we cannot fully distinguish from our model function with the existing dataset.  To estimate these errors, we fit the same data to a modified fit function that does not change sign for the two Dy isotopes, thus maximizing our sensitivity to any common-mode systematics, and replace our original statistical error bars on each coefficient with their magnitudes in the fit to the modified function, if they are greater.  If our original signal for $c_{T(Y-Z)}$ were not due to systematics, and thus truly due to Lorentz symmetry violation, we would expect to obtain a small value for $c_{T(Y-Z)}$ in the fit to the modified function.  Instead, we observe that the mean value of $c_{T(Y-Z)}$ without isotopic sign reversal is nearly as large as it is for the fit to the Lorentz-violating model, and thus conclude that this is not evidence for violation of LLI.  On the other hand, our estimates of the systematic error in our fits to $c_{TX}$ and $c_{T(Y+Z)}$ is comparatively quite low.  This is consistent with the hypothesis that our systematic background modulates with a period of one or half a day, and averages out over longer times, since our sensitivity to $c_{TX}$ and $c_{T(Y+Z)}$ comes primarily from the once-yearly modulated scalar term in Eq.~(3), while $c_{T(Y-Z)}$ is equally sensitive to both yearly and daily modulations proportional to the quadrupole term (see Table~\ref{tab:ctjcoefs}).

In both cases, the uncorrelated combinations of coefficients reported in Table~II were found by diagonalizing the covariance matrix from the least-squares fit to Eqs.~\eqref{eq:fitcjk} and \eqref{eq:fitctj}.

All frequency measurements are made relative to an HP5061A Cs frequency reference rated for a fractional-frequency stability better than $10^{-12}$ for $10^4$ s of averaging. The Cs reference is always compared to a GPS disciplined Symmetricom TS2700 Rb oscillator to verify that drifts of the reference frequency over the full course of the data collection period are below the $10^{-11}$ level. The results of this work rely on a fractional measurement precision of greater than $10^{-10}$, so instability of the frequency reference can be neglected in the analysis.  Though in general, violation of LLI may affect the frequency of this reference, such effects can be neglected here.  The relevant quadrupole term is strongly suppressed due to the symmetries of the $6s^{1}$ state~\cite{Kostelecky99a}, and the scalar shift is expected to be $\sim (10\text{ GHz})c_{\mu\nu}$~\cite{Hohensee2011}, while as Table~I shows, the splitting between the $A$ and $B$ states of Dy will be $\sim(100\text{ THz})c_{\mu\nu}$ or greater.


\begin{thebibliography}{999}

\frenchspacing

\bibitem{MTW} C.W. Misner, K.S. Thorne, and J.A. Wheeler, \emph{Gravitation} (Freeman, San Francisco, 1970).

\bibitem{Kostelecky:1989} V.A. Kosteleck\'y and S. Samuel, Phys. Rev. D {\bf 39,} 683 (1989).

\bibitem{Damour:1996} T. Damour, Classical  Quantum Gravity {\bf 13,} A33 (1996).

\bibitem{datatables} V.A. Kosteleck\'y and N. Russell, Rev. Mod. Phys. {\bf 83,} 11 (2011).

\bibitem{Will:2006} C. M. Will, Living Rev. Relativity {\bf 9,} 3 (2006).

\bibitem{Kostelecky:98} D. Colladay and V.A. Kosteleck\'{y}, Phys. Rev. D {\bf 55}, 6760 (1997); {\bf 58}, 116002 (1998);V.A. Kosteleck\'y, Phys. Rev. D {\bf 69,} 105009 (2004).

\bibitem{KosteleckyTasson2010} V.A. Kosteleck\'y and J.D. Tasson, Phys. Rev. D {\bf 83,} 016013 (2011).

\bibitem{Dzuba86} V.A. Dzuba, V.V. Flambaum, and I.B. Khriplovich,
%Enhancement of P- and T-nonconserving effects in rare-earth atoms,
Z. Phys. D: {\it Atoms, Molecules and Clusters}, {\bf 1}, 243-245 (1986).

\bibitem{Dzuba99ab} V.A. Dzuba, V.V. Flambaum, and J.K. Webb,
% Space-Time Variation of Physical Constants and Relativistic
% Corrections in Atoms,
Phys. Rev. Lett., {\bf 82}, 888 (1999);Phys. Rev. A, {\bf 59}, 230 (1999).

\bibitem{Dzuba94}   V.A. Dzuba, V.V. Flambaum, and M.G. Kozlov,
% PNC in Dy
Phys. Rev. A, {\bf 50}, 3812 (1994).

\bibitem{Dzuba03}  V.A. Dzuba, V.V. Flambaum, and M.V. Marchenko,
%Relativistic effects in Sr, Dy, YbII and YbIII and search for variation
%of the fine structure constant,
Phys. Rev. A, {\bf 68}, 022506 (2003).

\bibitem{Mueller2007} H. M\"uller \emph{et al.}, P.L. Stanwix, M.E. Tobar, E. Ivanov, P. Wolf, S. Herrmann, A. Senger, E.V. Kovalchuk and A. Peters, 
Phys. Rev. Lett. {\bf 99}, 050401 (2007).

\bibitem{Altschul:2010a} B. Altschul, Phys. Rev. D {\bf 82,} 016002 (2010).

\bibitem{Altschul:2006} B. Altschul, Phys. Rev. Lett. {\bf 96,} 201101 (2006); Phys. Rev. D {\bf 74,} 083003 (2006).

\bibitem{Vessot:1980} R.F.C. Vessot \emph{et al.}, Phys. Rev. Lett. {\bf 45,} 2081 (1980).

\bibitem{Hohensee2011} M.A. Hohensee, S. Chu, A. Peters, and H. M\"uller, Phys. Rev. Lett. {\bf 106,} 151102 (2011).


\bibitem{Kostelecky99a} V.A. Kosteleck\'{y} and C.D. Lane, Phys. Rev. D {\bf 60}, 116010 (1999).


\bibitem{Bluhm2003} R. Bluhm, V.A. Kosteleck\'{y}, C.D. Lane, and N. Russell, Phys. Rev. Lett., {\bf 88}, 090801 (2002); Phys. Rev. D {\bf 68}, 125008 (2003).

\bibitem{Altschul2010} B. Altschul, Phys. Rev. D {\bf 81}, 041701 (2010).

\bibitem{Kostelecky99b} V.A. Kosteleck\'{y} and C.D. Lane, J. Math. Phys. (N.Y.) {\bf 40}, 6245 (1999).

\bibitem{HohenseeMuellerToBePublished} M.A. Hohensee and H. M\"uller (to be published).

\bibitem{Dzuba08}  V.A. Dzuba and V.V. Flambaum,
%       Relativistic corrections to transition frequencies of
%       Ag~I, Dy~I, Ho~I, Yb~II, Yb~III, Au~I and Hg~II
%       and search for variation of the fine structure constant,
%       arXiv:0712.3621 (2007);
       Phys. Rev. A, {\bf 77}, 012514 (2008); Phys. Rev. A, {\bf 77}, 012515 (2008).
%\bibitem{Dzuba08a}  V.A. Dzuba and V.V. Flambaum,
%%       Relativistic corrections to transition frequencies of Fe~I and
%%       search for variation of the fine structure constant,
%       Phys. Rev. A, {\bf 77}, 012514 (2008).

\bibitem{VN} V.A. Dzuba,
%      $V^{N-M}$ approximation for atomic calculations,
      Phys. Rev. A, {\bf 71}, 032512 (2005);
V.A. Dzuba and V.V. Flambaum,
%      Core-valence correlations for atoms with open shells,
Phys. Rev. A. {\bf 75}, 052504 (2007).


\bibitem{Martin} W.C. Martin, R. Zalubas, and L. Hagan,
{\it Atomic Energy Levels - The Rare-Earth Elements}, NIST, Washington (1978).

\bibitem{Dzuba2010} V.A. Dzuba and V.V. Flambaum,
%      Core-valence correlations for atoms with open shells,
      Phys. Rev. A. {\bf 81}, 052515 (2010).




\bibitem{supplement} See Supplemental Material that follows for the derivation of the matrix elements of the states $A$ and $B$, and the expression for Eq.~\eqref{averes} in terms of $c_{\mu\nu}$ in the SCCEF.

\bibitem{Nguyen2004} A.T. Nguyen, D. Budker, S. K. Lamoreaux and J. R. Torgerson, Phys.\ Rev. A {\bf 69}, 022105 (2004).

\bibitem{Budker07} A. Cing$\ddot{o}$z, A. Lapierre, A.-T. Nguyen, N. Leefer,
D. Budker, S. K. Lamoreaux, and J. R. Torgerson,
%Title: Limit on the temporal variation of the fine-structure constant
%using atomic dysprosium
Phys.  Rev.  Lett. {\bf 98}, 040801 (2007);S.J. Ferrell, A. Cing$\ddot{o}$z, A. Lapierre, A.-T. Nguyen,
N. Leefer, D. Budker, V.V. Flambaum, S.K. Lamoreaux, and J.R. Torgerson,
Phys. Rev. A {\bf 76}, 062104 (2007).

\bibitem{Leefer2012} N. Leefer, C.T.M. Weber, A. Cing\"oz, J.R. Togerson, and D. Budker, arXiv:1304.6940 (2013).
%N.A. Leefer \emph{et al.}, to be published.



\bibitem{Budker08} N. Leefer, A. Cing$\ddot{o}$z, D. Budker,
  S.J. Ferrell, V.V. Yashchuk, A. Lapierre, A.-T. Nguyen,
  S.K. Lamoreaux, and J.R. Torgerson, in {\it Proceedings of the 7th
    Symposium Frequency Standards and Metrology, Asilomar, October
    2008}, edited by Lute Maleki, (World Scientific, Singapore, 2009), pp. 34-43.

\bibitem{LeeferBlackbody} C.T.M. Weber \emph{et al.}, to be published.


\bibitem{Herrmann2009Eisele2009} S. Herrmann, A. Senger, K. M\"ohle, M. Nagel, E.V. Kovalchuk and A. Peters, Phys. Rev. D. {\bf 80,} 105011 (2009); Ch. Eisele, A.Yu. Nevsky, and S. Schiller, Phys. Rev. Lett. {\bf 103,} 090401 (2009).

\bibitem{Klinkhamer:2008a} F.R. Klinkhamer and M. Schreck, Phys. Rev. D {\bf 78,} 085026 (2008).

\bibitem{Lev09} M. Lu, S.H. Youn, and B.L. Lev,
% arXiv:0912.0050 (2009).
%    Title: Trapping ultracold dysprosium: a highly magnetic gas for
%    dipolar physics
Phys.. Rev. Lett. {\bf 104}, 063001 (2010).

\bibitem{Szmytkowski2007} R. Szmytkowski, J. Math. Chem. {\bf 42}, 397 (2007).


\end{thebibliography}
\end{document}